\documentclass[prb,nofootinbib,twocolumn,superscriptaddress]{revtex4} 

\usepackage{graphicx}
\usepackage{dcolumn}
\usepackage{bm}
\usepackage{threeparttable}
\usepackage{times}
\usepackage{mathptmx}
\usepackage{lscape}
\usepackage{natbib}
\usepackage{amsmath}
\usepackage{amssymb}
\usepackage{braket}
\usepackage{comment}
\usepackage{color}

\def\degree{\kern-.2em\r{}\kern-.3em}

\begin{document}


\title{  Tropical Limit for Configurational Geometry in Discrete Thermodynamic Systems  }
   
\author{Koretaka Yuge}
\affiliation{
Department of Materials Science and Engineering,  Kyoto University, Sakyo, Kyoto 606-8501, Japan\\
}%

\author{Shouno Ohta}
\affiliation{
Department of Materials Science and Engineering,  Kyoto University, Sakyo, Kyoto 606-8501, Japan\\
}%

\begin{abstract}
{ 
For classical discrete systems with constant composition (typically referred to substitutional alloys) under thermodynamically equilibrium state, macroscopic structure should in principle depend on temperature and many-body interaction through Boltzmann factor, $\exp\left(-\beta E\right)$. 
Despite this fact, our recently find that (i) thermodynamic average for structure can be characterized by a set of special microscopic state whose structure is independent of energy and temperature,\cite{em0} and (ii) bidirectional-stability character for thermodynamic average between microscopic structure and potential energy surface is formulated without any information about temperature or many-body interaction.\cite{embd} 
These results strongly indicates the significant role of configurational geometry, where ``anharmonicity in structural degree of freedom (ASDF)'' that is a vector field on configuration space,  plays central role, intuitively corresponding to nonlinearity in thermodynamic average depending only on configurational geometry. 
Although ASDF can be practically drawn by performing numerical simulation based on such as Monte Carlo simulation, it is still unclear how its entire character is dominated by geometry of underlying lattice. 
We here show that by applying the limit in tropical geometry (i.e., tropical limit) with special scale-transformation  to discrete dynamical system for ASDF, we find tropical relationships between how nonlinearity in terms of configurational geometry expands or shrinks and geometric information about underlying lattice for binary system with a single structural degree of freedom (SDF). 
The proposed tropical limit will be powerful procedure to simplify analyzation of complicated nonlinear character for thermodynamic average with multiple SDF systems. 
  }
\end{abstract}


\maketitle

\section{Introduction}
When potential energy surface (PES) (or many-body interaction) is once given, structure (under prepared coordination $\left\{ q_{1},\cdots, q_{f} \right\}$) in thermodynamically equilibrium state with $f$-degree of freedoms in classical discrete system under constant composition (typically referred as substitutional crystalline solids) can be obtained through canonical average, whose summation is taken over all possible microscopic states on configuration space with weight of Boltzmann factor, 
$\exp\left( -\beta U \right)$.
Since number of microscopic states astronomically increases with increase of the system size and/or the number of components, theoretical techniques have been amply developed to effectively predict macroscopic properties including Metropolis algorism, entropic sampling and Wang-Landau sampling.\cite{mc1,mc2,mc3,wl} Althouth the current classical statistical mechamics successfully predict properties for given many-body interactions, it has been generally unclear how the canonical average can be (or cannot be) bidirectionally stable between PES and structure (i.e., stability from PES to structure, and from structure to PES).  

Very recently, we show that withouth using any information about temperature or PES, the bidirectional stability relationship (BSR) is quantitatively formulated by newly-introduced vector field on configuration space, ``anharmonicity in structural degree of freedom (ASDF)''\cite{asdf} depending only on configurational geometry \textit{before} applying many-body interaction to the system.\cite{embd} More specifically, hypervolume correspondence between Euclid space of structure $\left( q_{1},\cdots, q_{f} \right)$ and of PES $\left( \Braket{U|q_{1}}, \cdots, \Braket{U|q_{f}} \right)$ is characterized by sum of divergence and Jacobian of ASDF. 
Here, potential energy $U$ for any microscopic state $Q$ under complete orthonormal basis is exactly given by 
\begin{eqnarray}
U_{p}\left( Q \right) = \sum_{i=1}^{f} \Braket{U_{p}|q_{i}} q_{i}\left( Q \right),
\end{eqnarray}
where $\Braket{\quad|\quad}$ denotes inner product on configuration space, i.e., trace over all possible microscopic states, and subscript $p$ denotes the choice of landscape of the PES.
So far, although the ASDF intuitively corresponds to nonlinearity in the thermodynamic average in terms of configurational geometry, and can be estimated  as  time evolution of a special discrete dynamical system from numerical simulation, it is still unclear how its entire character is dominated by geometry of underlying lattice. 
Here we show that by applying limit in tropical geometry (i.e., tropical limit) to the dynamical system, we clarify how configurational nonlinearity is characterized by lattice geometry for equiatomic binary system. Its derivation and the detailed discussions are shown below.

\section{Derivation and Applications}
\subsection{Preparation for derivation}
Before formulating the nonlinearity in thermodynamic average in terms of lattice geometry, we first briefly explain the basic concept of the ASDF, BSR and related discrete dynamical system.
In the present study, we employ generalized Ising model\cite{gim} (GIM) providing a complete orthonormal basis function for any configuration on given lattice. For instance, in A-B binary system,  occupation of a lattice site $i$ is specified by spin variables, $\sigma_{i}=+1$ for A and
$\sigma_{i}=-1$ for B. Using the spin variables, basis function is given by
\begin{eqnarray}
\phi_{s}^{\left( d \right)} = \Braket{ \prod_{i\in s}\sigma_{i} }_{d}.
\end{eqnarray}
Here, $\Braket{\quad}_{d}$ means taking average for microscopic state $d$, taking over symmetry-equivalent figure $s$.
When structure $Q$ is interpreted as $f$-dimensional vector $Q=\left( q_{1},\cdots, q_{f} \right)$ under the orthonormal basis, ASDF is defined as the following vector field at $Q$:
\begin{eqnarray}
\label{eq:asdf}
A\left( Q \right) = \left\{ \phi_{\textrm{th}}\left( \beta \right)\circ \left( -\beta\cdot \Lambda \right)^{-1} \right\}\cdot Q - Q,
\end{eqnarray}
where $\phi_{\textrm{th}}$ denotes canonical average, $\beta$ inverse temperature, $\circ$ composite map, and $\Lambda$ is $f\times f$ real symmetric matrix of 
\begin{eqnarray}
\label{eq:lambda}
\Lambda_{ik} = \sqrt{\frac{\pi}{2}} \Braket{q_{k}}_{2}\Braket{q_{i}}_{k}^{\left( + \right)},
\end{eqnarray}
where $\Braket{\quad}_{2}$ denotes taking standard deviation for CDOS, and $\Braket{q}_{k}^{\left( + \right)}$ represents taking partial linear average satisfying $q\ge 0$. Note that here and hereinafter, structure $Q$ is measured from center of gravity of CDOS. Important points here are that (i) these average and standard deviation are taken \textit{before} applying many-body interaction to the system, and (ii) image of the composite map in 
Eq.\eqref{eq:lambda} is independent of temerature and of many-body interaction: Therefore, we can know vector field of ASDF without any information about temperature or energy, i.e., it can be \textit{a priori} known depending only on configurational geometry.
Since we have shown that $\Lambda$ is a \textit{unique} representation of linear part in canonical average in terms of configurational geometry,\cite{nln} ASDF in Eq.~\eqref{eq:asdf} 
is a measure of nonlinearity in $\phi_{\textrm{th}}$ defined on individual configuration $Q$. 

Next, we briefly describe the BSR in $\phi_{\textrm{th}}$. The BSR for structure Q has been defined as 
\begin{eqnarray}
B^{\left( Q \right)} = \log \left( \frac{R^{\left( Q \right)}}{ R^{\left( 0 \right)} } \right)
\end{eqnarray}
with 
\begin{eqnarray}
R^{\left( Q \right)} = \frac{\mathbf{dQ}^{\left( Q \right)}}{ \mathbf{dU}^{\left( Q \right)} },
\end{eqnarray}
where $\mathbf{dQ}^{\left( Q \right)}$ and $\mathbf{dU}^{\left( Q \right)}$ respectively denotes hypervolume in $f$-dimensional $Q$- and $PES$-space defined above, and superscript $\left( 0 \right)$ denotes structure at center of gravity of CDOS (i.e., disordered structure). Therefore, this definition of BSR intuitively provides that $B>0$ is $Q\mapsto U$ stable for $\phi_{\textrm{th}}$ as a map, $B<0$ is $U\mapsto Q$ stable, and $B=0$ is bidirectionally stable. 
We have shown that the BSR can be quantitatively formulated using only of the ASDF:\cite{embd}
\begin{eqnarray}
\label{eq:bsr}
B=\log \left|  1 + \mathrm{div} A + \sum_{F} J_{F}\left[ \frac{\partial A}{\partial Q }\right] + J\left[ \frac{\partial A}{\partial Q} \right]    \right|, 
\end{eqnarray}
where $J$ denotes Jacobian, and summation $F$ is taken over all possible subspaces derived from configuration space considered. 
From Eq.~\eqref{eq:bsr}, we can clearly see that the BSR can be characterized by information only about configurational geometry, and the BSR should be investigated through analyzing the trajectory of the following discrete dynamical system for ASDF:
\begin{eqnarray}
\label{eq:ds}
Q_{a+1} = Q_{a} + A\left( Q_{a} \right),
\end{eqnarray}
since we see the BSR for $Q_{a+1}$ at $Q_{a}$. Note that $A\left( Q_{a} \right)$ is a scalar for a considered single coordination. We have previously shown\cite{azero} that at center of gravity of CDOS, $\phi_{\textrm{th}}$ becomes a linear map, i.e., 
\begin{eqnarray}
\label{eq:azero}
A\left( Q_{a} = 0 \right) = 0.
\end{eqnarray} 

\subsection{Tropical limit of discrete dynamical system}
From above discussions, we can see that although BSR is characterized by ASDF, i.e., geometrical nonlinearity, it is still unclear how entire character of BSR is dominated by geometry of underlying lattice, since drawing vector field ASDF for large systems typically requires  numerical simulation clearly seen from 
Eq.~\eqref{eq:asdf}. 
To overcome this problem, some \textit{special} limit for ASDF would be naturally suggested, to focusing on relationships between lattice geometry and the nonlinearity.
Here, our strategy is to applying tropical limit to the discrete dynamical system in order to \textit{magnify} relationships to be interest. 
Briefly, tropical limit for multi-order real polynomials is interpreted as map from $\mathbb{R}_{>0}, +, \times$ to $\mathbb{R}, \otimes, \oplus$ thrugh the following: 
\begin{eqnarray}
\label{eq:trop}
&&\mathbb{R}_{>0} \ni x, y: \nonumber \\
&& x\mapsto t^{X},\quad y\mapsto t^{Y}\nonumber \\
&&X \oplus Y := \lim_{t\to\infty} \log_{t} \left( t^{X} + t^{Y} \right) = \max\left\{ X,Y \right\}\nonumber \\
&&X \otimes Y := \lim_{t\to\infty} \log_{t} \left( t^{X} \cdot t^{Y} \right) = X + Y,
\end{eqnarray}
where $\oplus$ and $\otimes$ respectively corresponds to tropical sum and tropical product, naturally inducing max-plus algebra. 
Note that the tropical limit has been naturally adopted in statistical mechanics, e.g., free energy $F=-kT\sum_{E} \exp\left( -\beta E \right)$ with $\beta\to \infty$ induces tropical limit, where $F$ merely equals to minimum energy.\cite{trop} However, we emphasize here that the present tropical limit is completely different from such adoption.
Because of the above character, we cannot simply apply tropical limit to the discrete dynamical system because (i) domain of $Q$ includes negative value, and (ii) $A\left( Q \right)$ can take positive, zero or negative sign, depending on $Q$. Problem (i) can be straightforwardly solved by variable transformation of $Q' = Q + B$ where e.g., $B\ge 1$ such that range of $Q'$ takes positive sign. 
Therefore, nontrivial problem is (ii). 
\begin{figure}[h]
\begin{flushright}
\includegraphics[width=0.84\linewidth]{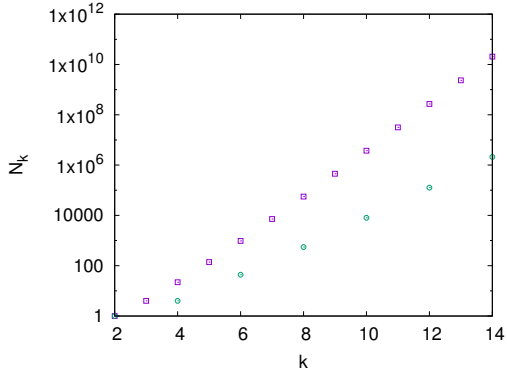}
\vspace{7mm} 
\caption{ $N_{k}$ as a function of order $k$ for fcc 1-NN (open squares) and 2-NN (open circles) pair.  }
\label{fig:sc}
\end{flushright}
\end{figure}
One straightforward approach for the problem (ii) is to take square for both sides of Eq.~\eqref{eq:ds} after the variable transformation, leading to
\begin{eqnarray}
\label{eq:ds0}
\left( Q'_{a+1} \right)^{2} + \left( Q'_{a} \right)^{2} = 2Q'_{a+1}Q'_{a} + A^{2}.
\end{eqnarray}
Then corresponding tropical limit becomes
\begin{eqnarray}
\max\left\{ 2\Omega_{a+1}, 2\Omega_{a} \right\} = \max\left\{ \Omega_{a+1} + \Omega_{a}, 0 \right\},
\end{eqnarray}
which has solution of
\begin{eqnarray}
\label{eq:sol0}
\Omega_{a+1} = \Omega_{a} \quad \left( \Omega_{a} \ge 0 \right) \nonumber \\
\Omega_{a+1} = 0 \quad \left( \Omega_{a} < 0 \right).
\end{eqnarray}
However, above dynamical system of Eq.~\eqref{eq:ds0} has several essential problems: Taking square of both sides can lose part of information about original dynamical system of Eq.~\eqref{eq:ds}, resultant solution of Eq.~\eqref{eq:sol0} exhibit completely different behavior of practical vector field $A$ (e.g., fixed point $\Omega_{a+1} = \Omega_{a}$ is limited to finite section for any given lattice), there exists no information about geometry of underlying lattice, and origin of $\Omega=0$ is not well defined (independent of parameter $B$). 
These facts strongly suggest that do not take square of both sides, and  $A$ should not be treated merely as scalar since $A$ is a function both of $Q_{a}$ and $B$. 
Therefore, to explicitly apply tropical limit for ASDF, $A$ should be represented by polynomial in terms of $Q_{a}$ and $B$. To achieve this, we start from series expansion of $A$ in terms of moments of CDOS, given by
\begin{widetext}
\begin{eqnarray}
A = \sum_{k=2}^{\infty} \frac{1}{k!} \frac{1}{\Braket{Q^{2}}^k } \left\{ \Braket{Q^{\left( k+1 \right)}} + \sum_{j} c_{j}\cdot \prod_{\left\{ w_{j} | \sum w_{j} = \left( k+1 \right)  \right\}}\Braket{Q^{w_{j}}} \right\}\cdot Q^{k}, \nonumber \\
\quad
\end{eqnarray}
\end{widetext}
where $\Braket{\quad}$ denotes taking linear average on CDOS, and $c_{j}$ corresponds to coefficient for the expansion. 
We here focus on a single $Q$ as pair correlation on equicompositional binary system, where only diagrams for moments, with number of odd-time occupation by lattice points taking zero, dominantly contribute to $A$ at thermodynamic limit. 
Under this condition, we here simplify expression of $A$ as
\begin{eqnarray}
\label{eq:aj}
A\simeq \sum_{k=2}^{\infty} \left[ \left( k+1 \right) \frac{J_{k+1}}{D} \sum_{m=0}^{k} \left\{ {}_{k}C_{m} \left( Q' \right)^{m} \left( -B \right)^{\left( k-m \right)} \right\} \right],
\end{eqnarray}
where $J_{k+1}$ denotes number of closed cycles per site consisting only of considered pair type, and $D$ number of pair per site. 
Note that from Eq.~\eqref{eq:aj}, since parity of maximum power of $e^{M}$ and that of maximum power of $Q'$ differs, hereinafter we naturally consider the condition that parity of $n$ and $\lambda$ differs.
Since number of simple cycle cannot be simply formulated in terms of lattice geometry, we here perform numerical simulation to \textit{exactly} count $N_{k}= kJ_{k}/D$ up to $k=14$ for e.g., fcc with 1 nearest-neighbor (1-NN) and 2-NN pair, shown in Fig.~\ref{fig:sc}. 
We can clearly see that $N_{k}\simeq e^{Mk}$ where $M$ depends on pair type except for $N_{k}=0$ with $k$ taking odd number in 2-NN pair. 
Hereinafter, we focus on the case of $N_{k}\simeq e^{Mk}$, while the following derivation can be straightforwardly extended to other cases including 2-NN pair.
Then coefficient for $\left( Q' \right)^{g}$ in Eq.~\eqref{eq:aj} can be geometric sequence with geometric ratio of $\sim \left|e^{M}\cdot\left( -B \right)\right| > 1$ for large $k$, 
\begin{eqnarray}
\label{eq:asim}
A\simeq \frac{1+e^{M}B}{e^{M}B} \lim_{n\to\infty}\left\{ e^{\left(n+1\right)M} \sum_{g=0}^{\lambda}{}_{n}C_{g}\left(-B\right)^{n-g}\left(Q'\right)^{g} \right\}
\end{eqnarray}
This transformation indicates that we naturally include coefficient for $Q'^{m}$ contributed from all possible $k\ge m$ in Eq.~\eqref{eq:aj}, while we terminate maximum power of $Q$ up to condition of $\lambda \ll k_{\textrm{max}} $ so that 
\begin{eqnarray}
\lim_{k_{\textrm{max}}\to\infty} \frac{ {}_{k_{\textrm{max}}}C_{\lambda} }{e^{k_{\textrm{max}}M}} = 0
\end{eqnarray}
is satisfied, where we take $n=k_{\textrm{max}}$.
Under this condition, dynamical system of Eq.~\eqref{eq:ds} for taking tropical limit can be transformed into
\begin{widetext}
\begin{eqnarray}
\label{eq:dsf}
Q'_{a+1} = Q'_{a} +\left( \frac{1+e^{2M}B}{e^{M}B}\right)e^{nM}\frac{\left( -B\right)^{n} \left\{ 1-\left(-B^{-1}Q'_{a}\right)^{\lambda+1} \right\}  }{ 1+B^{-1}Q'_{a}}.
\end{eqnarray}
\end{widetext}

It is clear that Eq.~\eqref{eq:dsf} should be analyzed individually by conditions of (i) parity of $n$, and (ii) whether or not $\left|\left( -B^{-1} \right)Q'\right|$ exceeds 1. 
We first start from the condition that $n$ is even with $\left|\left( -B^{-1} \right)Q'\right| < 1$. 
In this case, we have
\begin{eqnarray}
\label{eq:a01}
Q'_{a+1}Q'_{a}  + Q'_{a+1} = \left(Q'_{a}\right)^{2} + Q'_{a} + He^{nM},
\end{eqnarray}
where 
\begin{eqnarray}
H = \frac{1+e^{2M}B}{e^{M}B}.
\end{eqnarray}
We note here that by simply applying tropical limit to Eq.~\eqref{eq:a01}, geometric information of lattice is clearly dissapeared. 
Since we take $n\to\infty$, we can avoid this problem by explicitly consider $e^{n}$ as additional variable, where
\begin{eqnarray}
\label{eq:g}
\log_{t}e^{n} = \gamma.
\end{eqnarray}
Necessary condition for $\gamma$ will be naturally determined from physical insight in the following discussions. Substituting Eq.~\eqref{eq:g} and setting artificial translation for $Q$ of $B=1$, we obtain tropical limit for Eq.~\eqref{eq:a01} as

\begin{widetext}
\begin{eqnarray}
\label{eq:at01}
\max\left\{ \Omega_{a+1} + \Omega_{a}, \Omega_{a+1} \right\} = \max\left\{ 2\Omega_{a} , \Omega_{a} , M\gamma\right\}\quad \left( n:\textrm{even}, \Omega_{a} < 0 \right).
\end{eqnarray}
\end{widetext}
Since we here consider the case of $\left|\left( -B \right)Q'\right| < 1$ with $B=1$, Eq.~\eqref{eq:at01} corresponds to tropical limit of dynamical system under even number of $n$ with $\Omega_{a} < 0$ (because of $Q_{a}<1$, i.e., $Q_{a}$ below center of gravity of CDOS).
In a similar fashion, tropical limit for the dynamical system under other conditions with the same transformation of Eq.~\eqref{eq:g} are respectively given by
\begin{widetext}
\begin{eqnarray}
\label{eq:at02}
&&\max\left\{ \Omega_{a+1}+\Omega_{a}, \Omega_{a+1} \right\} = \max\left\{ 2\Omega_{a}, \Omega_{a}, \left(\lambda+1\right)\Omega_{a} + M\gamma, M\gamma \right\} \quad \left( n:\textrm{even}, \Omega_{a} \ge 0 \right) \nonumber \\
&& \max\left\{ \Omega_{a+1}+\Omega_{a}, \Omega_{a+1}, M\gamma \right\} = \max\left\{ 2\Omega_{a}, \Omega_{a} \right\} \quad \left( n:\textrm{odd}, \Omega_{a} < 0 \right) \nonumber \\
&& \max\left\{ \Omega_{a+1} + \Omega_{a}, \Omega_{a+1} + M\gamma \right\} = \max\left\{ 2\Omega_{a}, \Omega_{a},\left(\lambda+1\right)\Omega_{a}+M\gamma \right\} \quad \left( n:\textrm{odd}, \Omega_{a} \ge 0 \right). 
\end{eqnarray}
\end{widetext}
These derived tropical limit of dynamical system can exhibit different behavior w.r.t. sign of $M\gamma$. 
When $M\gamma$ takes positive value, we can clearly see for the cases of $\Omega_{a} < 0$ with $n$ taking even or odd that at 
$\Omega_{a+1}=M\gamma$ at $\Omega_{a}=0$: This means that at center of gravity of CDOS, $\phi_{\textrm{th}}$ is not a linear map since 
$\Omega_{a+1}-\Omega_{a}=M\gamma$, which loses important information about original dynamical system of Eq.~\eqref{eq:azero}. The exception is $M\gamma=0$, corresponding to $\gamma=0$. However, $\gamma=0$ should not be allowed since in this case, we get from Eq.~\eqref{eq:g} that 
\begin{eqnarray}
e^{n} = 1
\end{eqnarray}
regardless of the value of $t$. 
Therefore, in order to keep the dynamical system linear at center of gravity of CDOS, Eq.~\eqref{eq:even} should not be allowed. 
Under this condition, Eqs.~\eqref{eq:at01} and~\eqref{eq:at02} having solution lead to
\begin{widetext}
\begin{eqnarray}
\label{eq:final}
\Omega_{a+1} = \begin{cases}
    M\gamma & \left( \Omega_{a} \le M\gamma \right) \\
    \Omega_{a} & \left( M\gamma < \Omega_{a} < -\dfrac{M\gamma}{\lambda-1} \right) \\
    \lambda\Omega_{a} + M\gamma & \left( -\dfrac{M\gamma}{\lambda-1} \le \Omega_{a} \right)
  \end{cases}
\end{eqnarray}
\end{widetext}
We can now clearly see that $\Omega_{a+1}=\Omega_{a}$ at $\Omega_{a}=0$ is satisfied. Therefore, to  reasonablly keep $\phi_{\textrm{th}}$ as linear map at center of gravity of CDOS, $M\gamma < 0$ should be required. 
\begin{figure}
\begin{flushright}
\includegraphics[width=0.86\linewidth]{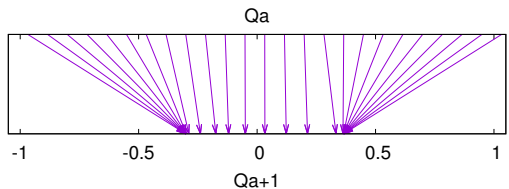}
\caption{Simulated discrete dynamical system of Eq.~\eqref{eq:ds} for fcc equiatomic binary system with 1-NN pair correlation. }
\label{fig:vec}
\end{flushright}
\end{figure}
\begin{figure}
\begin{center}
\includegraphics[width=0.8\linewidth]{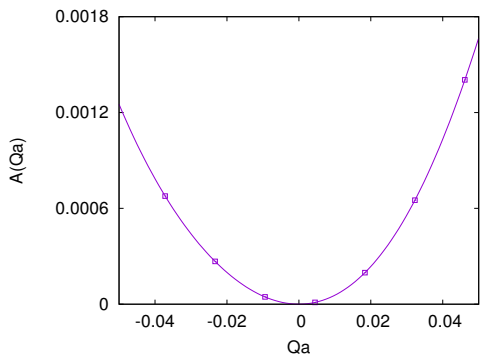}
\vspace{5mm} 
\caption{Value of ASDF near center of gravity of CDOS. }
\label{fig:lin}
\end{center}
\end{figure}
We first note that monotonic increase behavior for $\Omega_{a} \ge -M\gamma/\left(\lambda-1\right) > 0$, which has a deviation from original dynamical system, can be partly due to the approximation of $A$ with information about multi-order simple cycles in Eq.~\eqref{eq:asim}, and to the fact that concept of phase separation (i.e., vertext of configurational polyhedra for $Q>0$) can only be achieve at thermodynamic limit of $N\to\infty$, where multiple configuration with infinitesimal differences in correlations and in energy can exist near the vertext, which should be carefully investigated in future. 
Aside from this exception, dynamical system of Eq.~\eqref{eq:final} exhibit several important character of original dynamical system: (i) At $\Omega_{a} = M\gamma < 0$, divergence of the tropical vector field $\Omega$ becomes singular, which agrees with vertex of configurational polyhedra at negative sign of $Q_{a}$ behaves as adsorption point as shown in Fig.~\ref{fig:vec}, and (ii) linear region is asymmetric (linearity in $\phi_{\textrm{th}}$ is stronger for $Q_{a}< 0$ than that for $Q_{a}>0$, which also capture the asymmetric nonlinearity magnitude around center of gravity of CDOS as shown in Fig.~\ref{fig:lin}. Especially, character of (i) should be exact only at $\beta\to\infty$ where ASDF is not well-defined for original dynamical system. 
Therefore, it can be indicated that linearity in $\phi_{\textrm{th}}$ at disordered structure and singularity at ordered structure is dual, which should be further investigated in our future study.
Additional important point for Eq.~\eqref{eq:final} is that region for $\phi_{\textrm{th}}$ as linear map with increase of $M$: This can be intuitively interpreted that at limit of reducing spatial-constraint to constituents (by lattice) becoming free (i.e., continuous space with no constraint of lattice), linear map region naturally increases due to increase of $M$, which is consistent that deviation in landscape of CDOS from gaussian can typically reduces with reducing the spatial constraint, where we have shown that $\phi_{\textrm{th}}$ exactly becomes linear map if and only if CDOS takes Gaussian. 

\section{Conclusions}
By introducing special scale transformation, we construct tropical limit of dynamical system for representing nonlinearity in thermodynamic average coming from configurational geometry, for classical discrete systems. 
The tropical dynamical system explicitly reveals how entire character of vector field of anharmonicity in the structural degree of freedom (corresponding to geometrical nonlinearity) is dominated by geometry of underlying lattice.

\section{Acknowledgement}
This work was supported by Grant-in-Aids for Scientific Research on Innovative Areas on High Entropy Alloys through the grant number JP18H05453 and a Grant-in-Aid for Scientific Research (16K06704) from the MEXT of Japan, Research Grant from Hitachi Metals$\cdot$Materials Science Foundation.

\end{document}